\documentclass[12pt,a4paper]{article}
\pdfoutput=1
\usepackage{jcappub}
\usepackage{graphicx,amssymb,amsmath,xcolor}
\usepackage[normalem]{ulem}
\graphicspath{{figs/}}

\newcommand{\figw}[2]{\includegraphics[width=#1\textwidth]{#2}}

\newcommand{\black}{\color{black}}

\def\lsim{\raise0.3ex\hbox{$\;<$\kern-0.75em\raise-1.1ex\hbox{$\sim\;$}}}
\def\gsim{\raise0.3ex\hbox{$\;>$\kern-0.75em\raise-1.1ex\hbox{$\sim\;$}}}

\begin{document}

\begin{flushright}
INR-TH-2019-009
\end{flushright}

\title{%
	  Dark matter component decaying after recombination: constraints from diffuse gamma-ray and neutrino flux measurements
  }

\author[a,b]{O.~Kalashev,}
\author[a]{ M.~Kuznetsov,}
\author[a]{Y.~Zhezher}

\affiliation[a]{Institute for Nuclear Research of the Russian Academy of
	Sciences, Moscow, 117312, Russia}
\affiliation[b]{Moscow Institute for Physics and Technology, 9 Institutskiy per., Dolgoprudny, Moscow Region, 141701 Russia}

\abstract{
	We consider scenario of the dark matter consisting of two fractions, stable part being dominant and a smaller unstable fraction, which has decayed after the recombination epoch. It has been suggested in Ref.~\cite{Berezhiani:2015yta} that the above scenario may alleviate tension between high-redshift (CMB anisotropy) and low-redshift (cepheid variables and SNe Ia, cluster counts) cosmological measurements. We derive constraints on the heavy relics branching to $q\bar{q}$, $e^+e^-$, $\mu^+\mu^-$, $\tau^+\tau^-$, $\nu_e\bar{\nu_e}$, $\nu_\mu\bar{\nu_\mu}$,  $W^+W^-$ and $\gamma\gamma$ in the above scenario by comparison of the secondary $\gamma$ and $\nu$ fluxes produced by the process with recent diffuse $\gamma$ and $\nu$ flux measurements.
}
\emailAdd{kalashev@inr.ac.ru}
\emailAdd{mkuzn@inr.ac.ru}
\emailAdd{zhezher.yana@physics.msu.ru}

\maketitle

\section{Introduction}

New era in precise  determination  of  cosmological
parameters was opened by the WMAP~\cite{Hinshaw:2012aka} and Planck~\cite{Ade:2013zuv} measurements of the cosmic microwave background fluctuations. Surprisingly, it revealed the tension between the CMB based determination of the Hubble constant and the previous interpretation of $h$ value from direct low redshift measurements. Low-redshift determination of the Hubble constant, derived from the the Hubble Space Telescope (HST) cepheid and SNe Ia data lead to $h = 0.738 \pm 0.024$~\cite{Riess:2011yx}, while $h$ deduced from the Planck data is equal to $h = 0.6727 \pm 0.0066$~\cite{Ade:2013zuv}, showing $2.5\sigma$  discrepancy between two measurements. Other than that, such cosmological parameters as initial density perturbations $\sigma_8$ and mass density parameter $\Omega_m$ also show discrepancy for high-redshift measurements from  the  CMB and for low-redshift measurements from clusters as cosmological probes~\cite{Ade:2015fva}.

It was recently proposed~\cite{Berezhiani:2015yta} that this tension may be resolved if a certain fraction of the dark matter is unstable. In this model, it is suggested that the dark matter consists of two fractions: stable and unstable. The stable fraction is dominant, and only a small fraction of dark matter decays between recombination and the present epoch. In the follow-up papers~\cite{Chudaykin:2016yfk, Chudaykin:2017ptd} lensing constraints along with  baryon acoustic oscillation and redshift space distortions measurements have been used to limit the allowed range of decaying dark matter (DDM) fraction to $F < 4-7\ \%$ at $95\%\ \mbox{C.L.}$.

The DDM scenario finds another application in the possible explanation of the astrophysical neutrino flux discovered by the IceCube Collaboration~\cite{Aartsen:2013jdh,Aartsen:2017mau}. The distribution of IceCube events is isotropic in the sky, which together with Fermi data on the accompanying $\gamma$-ray flux makes it challenging to model the production of IceCube neutrinos in astrophysical sources~\cite{Murase:2013rfa} or in the present decay of DM~\cite{Kuznetsov:2016fjt, Kachelriess:2018rty}. \black In~\cite{Anchordoqui:2015lqa}, the DDM model was considered with the  dominant decay mode into the visible sector $X \rightarrow \nu \bar{\nu}$ with branching ratio $\mathbb{B}_{X\rightarrow \nu\bar{\nu}}\simeq 3\times 10^{-9}$.

The present study is aimed to derive the constraints on the heavy relics branching to a number of decay channels: $q\bar{q}$, $e^+e^-$, $\mu^+\mu^-$, $\tau^+\tau^-$, $\nu_e\bar{\nu_e}$, $\nu_\mu\bar{\nu_\mu}$,  $W^+W^-$ and $\gamma\gamma$, assuming scenario of Ref.~\cite{Berezhiani:2015yta}. While secondary $\nu$ lose their energy essentially through redshift only, secondary $e^+/e^-$ and $\gamma$-rays initiate electron-photon cascades on the CMB during their propagation. The cascade develops through the chain of inverse Compton scattering of electrons and $e^+e^-$ pair production by photons on CMB until the threshold for the pair production is achieved, below which the energy is collected in the form of effectively sterile photons~\cite{Berezinsky:2016feh}. We obtain the desired constraints by comparison of $\nu$- and $\gamma$-fluxes with observations. Namely, we compare the model gamma-ray flux with the Fermi LAT~\cite{Ackermann:2014usa} and EGRET~\cite{Strong:2004ry} isotropic diffuse $\gamma$-ray background (IGRB) estimates, while the model neutrino flux is compared with the neutrino flux upper-limits set by Super-Kamiokande~\cite{Zhang:2013tua, Richard:2015aua} and IceCube~\cite{Aartsen:2015xup, Aartsen:2017mau, Aartsen:2018vtx}.

Below, in Section~\ref{sec:analysis}, we describe in details the method we use to derive the constraints and present results along with discussion in Section~\ref{sec:results}.

\section{Analysis}\label{sec:analysis}
We consider the range of dark matter masses $600~{\rm GeV} \lesssim M_{DM} \lesssim 10^{15}$~GeV. 
For lower DM masses the primary photon energy is typically below the threshold for $e^+e^-$ pair production $E_{th}=m_e^2/E_{cmb}\simeq 300\mbox{GeV}$ at redshift $z=10^3$ and therefore the corresponding limits could be derived without taking into account the EM cascade development.
For $M_X/2\leq 100$TeV we use PPPC 4 DM ID toolkit~\cite{Cirelli:2010xx} to calculate the spectra of $e^+$, $e^-$, $\gamma$ and $\nu$ products of DM decay for several decay channels: $q\bar{q}$, $e^+e^-$, $\mu^+\mu^-$, $\tau^+\tau^-$, $\nu_e\bar{\nu_e}$, $\nu_\mu\bar{\nu_\mu}$,  $W^+W^-$ and $\gamma\gamma$.
For DM masses larger than 200~TeV we use the injection spectra of $\gamma$, $e^+$, $e^-$ and $\nu$ calculated for two benchmark decay channels: $q\bar{q}$ and $\nu\bar{\nu}$ with the numerical codes of Ref.~\cite{Aloisio:2003xj} and Ref.~\cite{Kachelriess:2018rty}, respectively~\footnote{As it was discussed in Ref.~\cite{Kachelriess:2018rty}, among all possible decay channels the $q\bar{q}$ and $\nu\bar{\nu}$ channels yield the softest and the hardest energy spectra, respectively, for both gamma-rays and neutrinos in the final state. Thus the constraints on HDM parameters that could be obtained for other decay channels or for their combinations should lie somewhere in between the constraints derived for the two aforementioned decay channels. In this sense we call $q\bar{q}$ and $\nu\bar{\nu}$ the benchmark channels.}. By $q\bar{q}$ and $\nu\bar{\nu}$ we mean the DM decay into quarks and neutrinos with uniform distribution in flavors.

We expect at least for $\gamma$-ray constraints to weaken with shorter DM decay time, since in this particular case the EM cascade develops on average in more energetic background and resulting diffuse $\gamma$-ray background is shifted towards lower energies where it is less constrained. Therefore to build conservative constraints we assume below that DM decays in time $\tau \propto H^{-1}$ 

Having $e^+/e^-$ and $\gamma$ spectra from DM decay we propagate them from $z_{\rm CMB}=1090$~\cite{Tanabashi:2018oca} using the TransportCR code~\cite{Kalashev:1999ma, Kalashev:2014xna}, developed for the simulation of ultra-high-energy cosmic rays and electron-photon cascade attenuation. The electron-photon cascade development essentially stops when photon energies reach the threshold $E_{th}$ for $e^+e^-$ pair production on CMB and afterwards the spectrum is only affected by the adiabatic Universe expansion. Due to the EM cascade universality~\cite{Berezinsky:2016feh} the final shape of $\gamma$-ray spectrum predicted is roughly the same for all the models considered in which the average initial electron and photon energy is well above $E_{th}$. Therefore for these models only total energy density coming to EM cascade is relevant for the constraint derivation.

\begin{figure}[!ht]
	\centering
	\figw{.49}{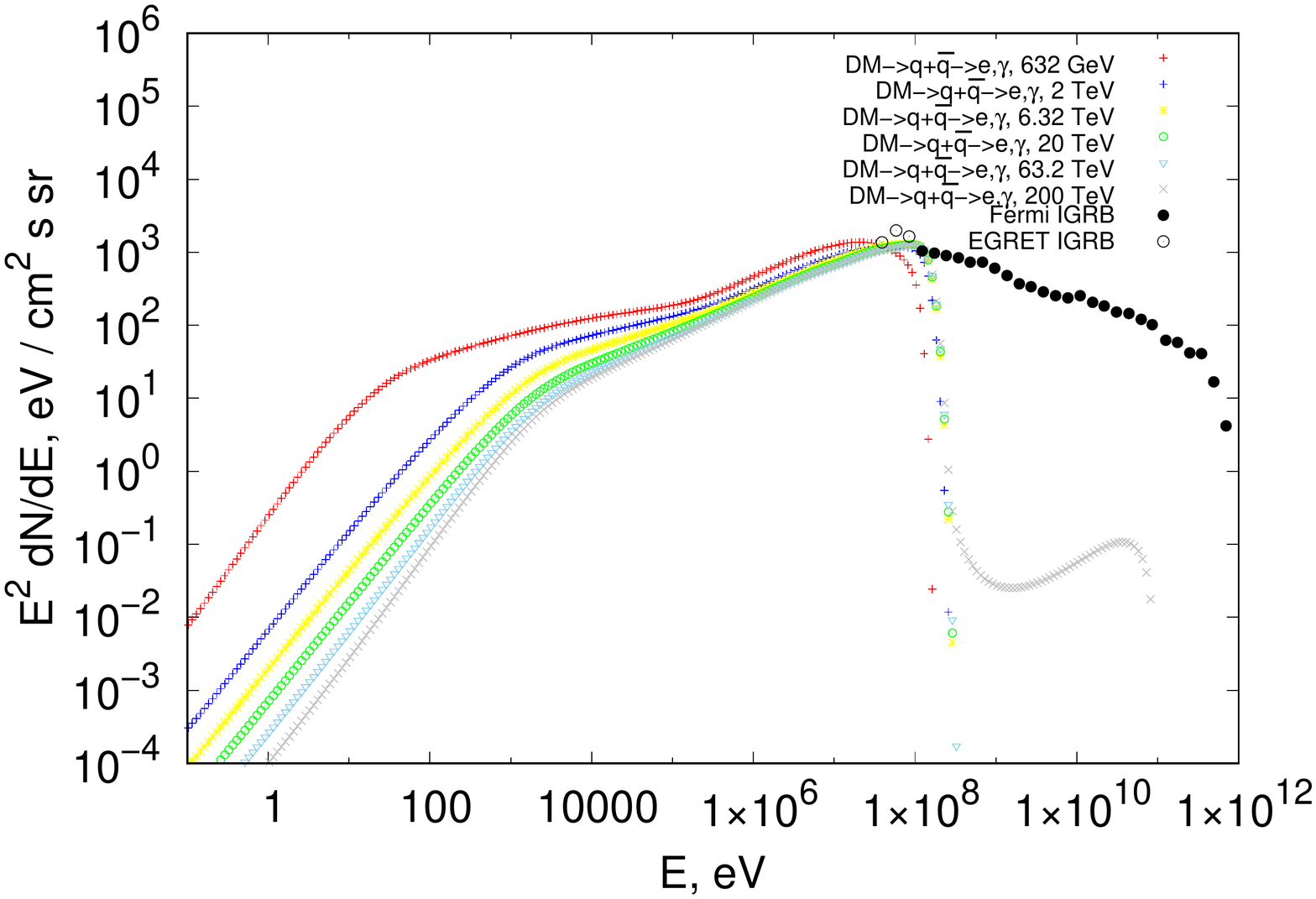}
	\figw{.49}{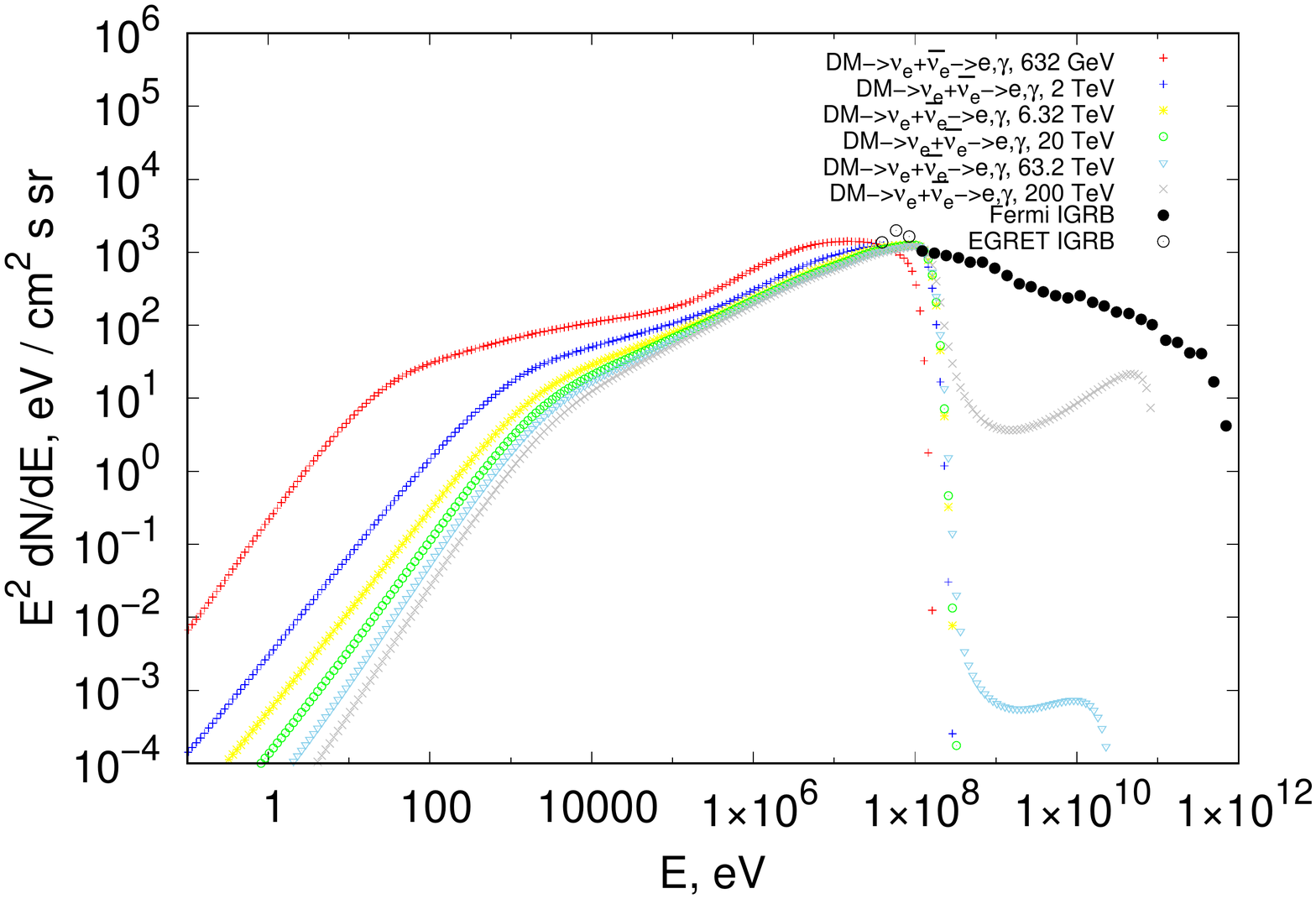}
	\caption{The spectra of secondary $\gamma$ from
	$q\bar{q}$ (left panel) and $\nu_e\bar{\nu_e}$ (right panel) decay for several DM particle masses, compared with Fermi LAT~\cite{Ackermann:2014usa} and EGRET~\cite{Strong:2004ry} IGRB estimates.}
	\label{fig:spectrum}
\end{figure}

Secondary $\gamma$-ray spectra, predicted by the model at redshift $z=0$, shouldn't contradict to the current IGRB estimates. In the present analysis, we have adopted the Fermi LAT IGRB data~\cite{Ackermann:2014usa}, derived in the energy range from $100\ \mbox{MeV}$ to $820\ \mbox{GeV}$ 
and estimates by EGRET~\cite{Strong:2004ry} derived in the energy range $30\ \mbox{MeV} -50\ \mbox{GeV}$. Overall IGRB spectrum considered is shown in Fig.~\ref{fig:spectrum}.

\begin{figure}[!ht]
	\centering
	\figw{.49}{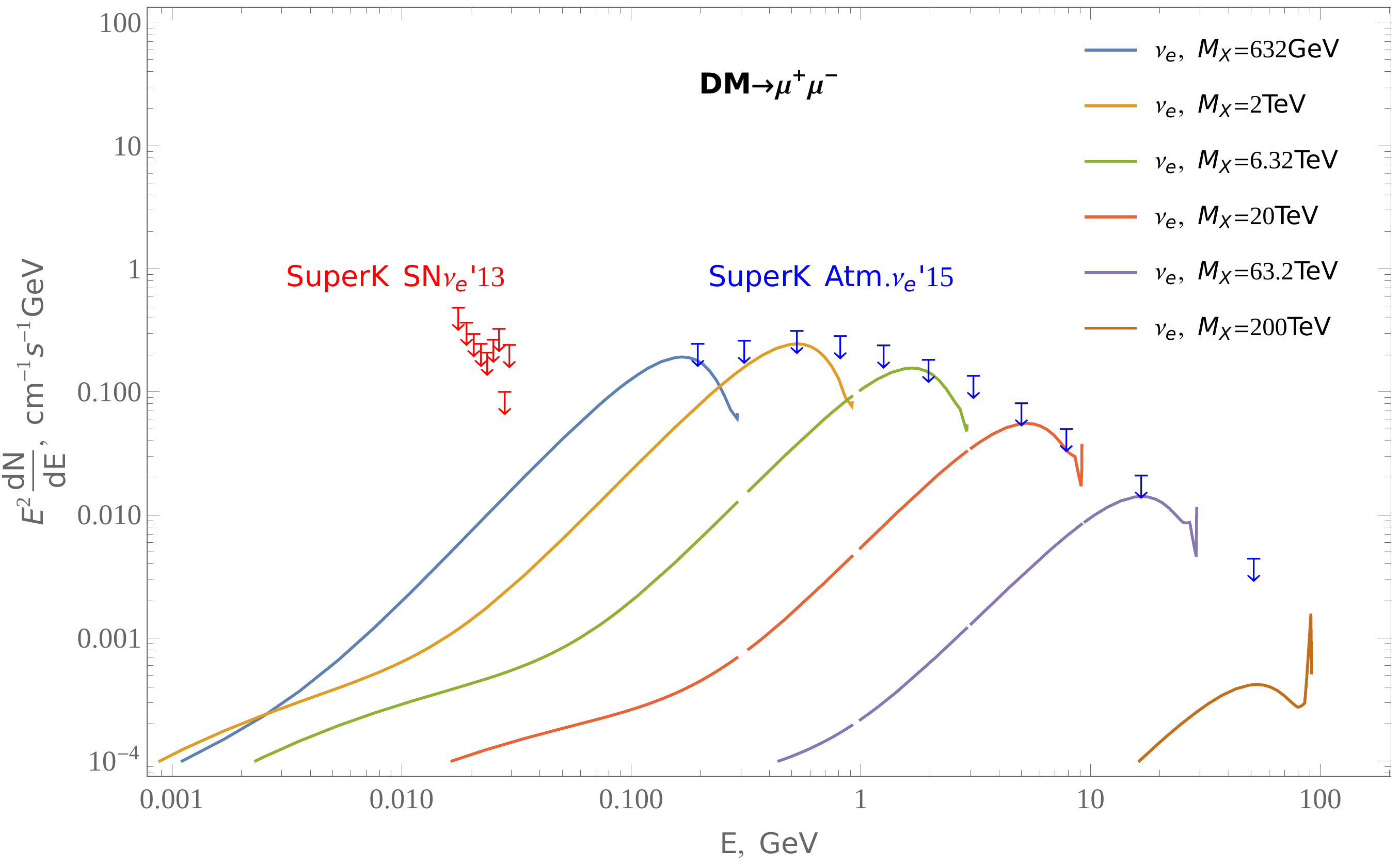}
	\figw{.49}{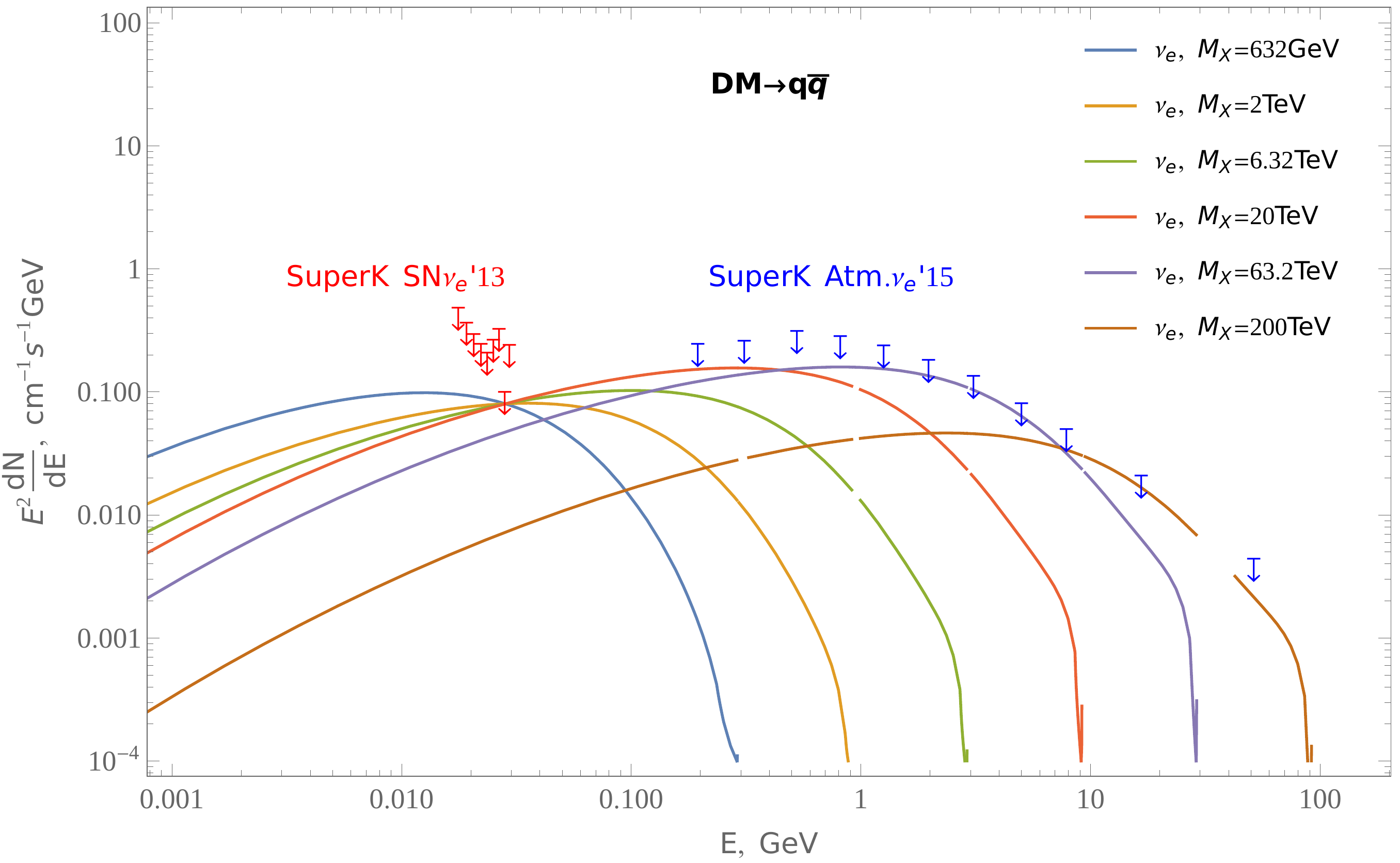}
	\figw{.49}{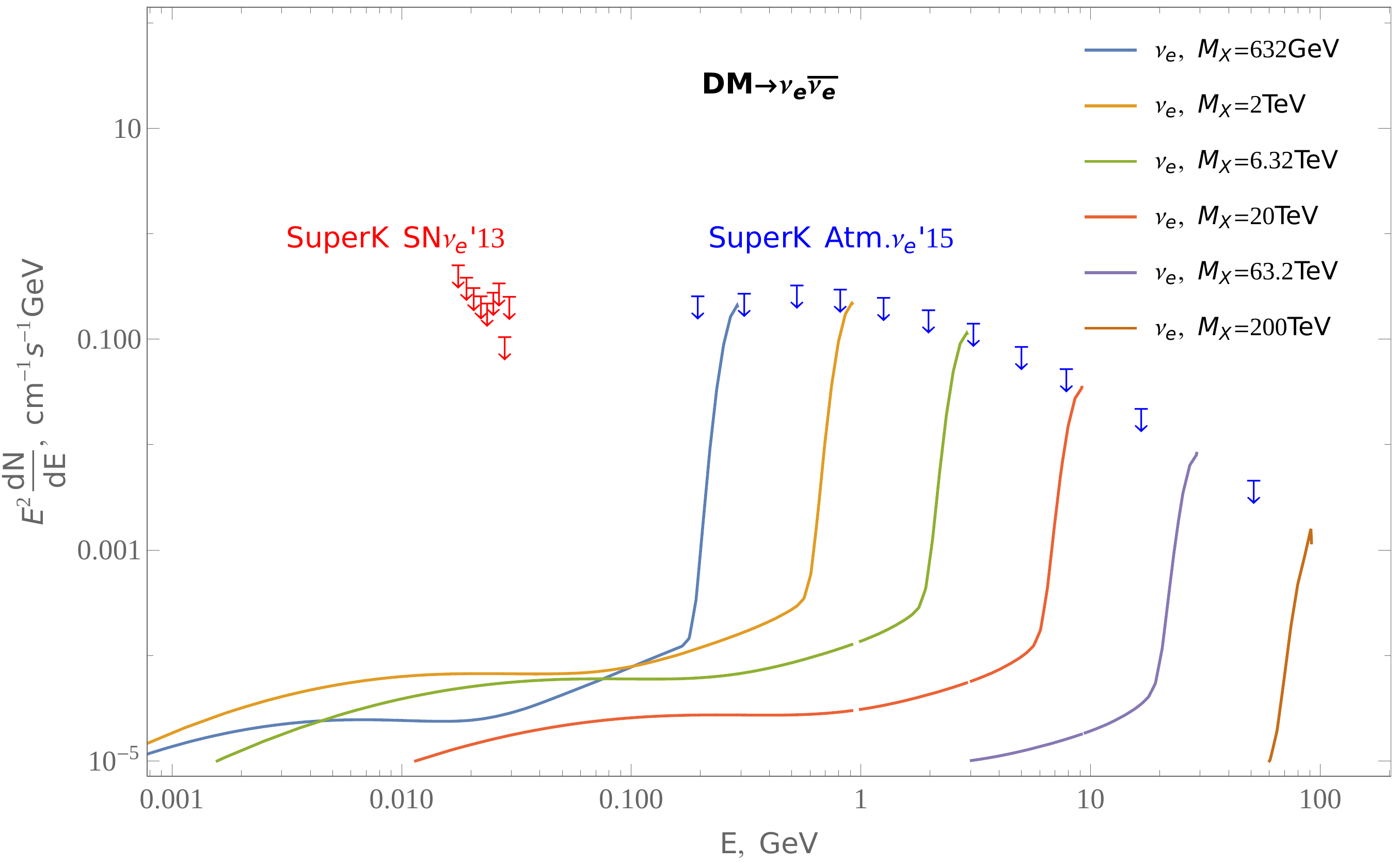}
	\caption{The spectra of $\nu_e$ from $\mu^+\mu^-$, $q\bar{q}$ and $\nu_e\bar{\nu_e}$ decay (from top to bottom) for several DM particle masses,	compared with Super-Kamiokande limits on diffuse electronic neutrino flux from supernovae~\cite{Zhang:2013tua} and atmospheric electronic neutrino flux limits~\cite{Richard:2015aua}.}
	\label{fig:spectrum_nu_low}
\end{figure}

\begin{figure}[!ht]
	\centering
	\figw{.49}{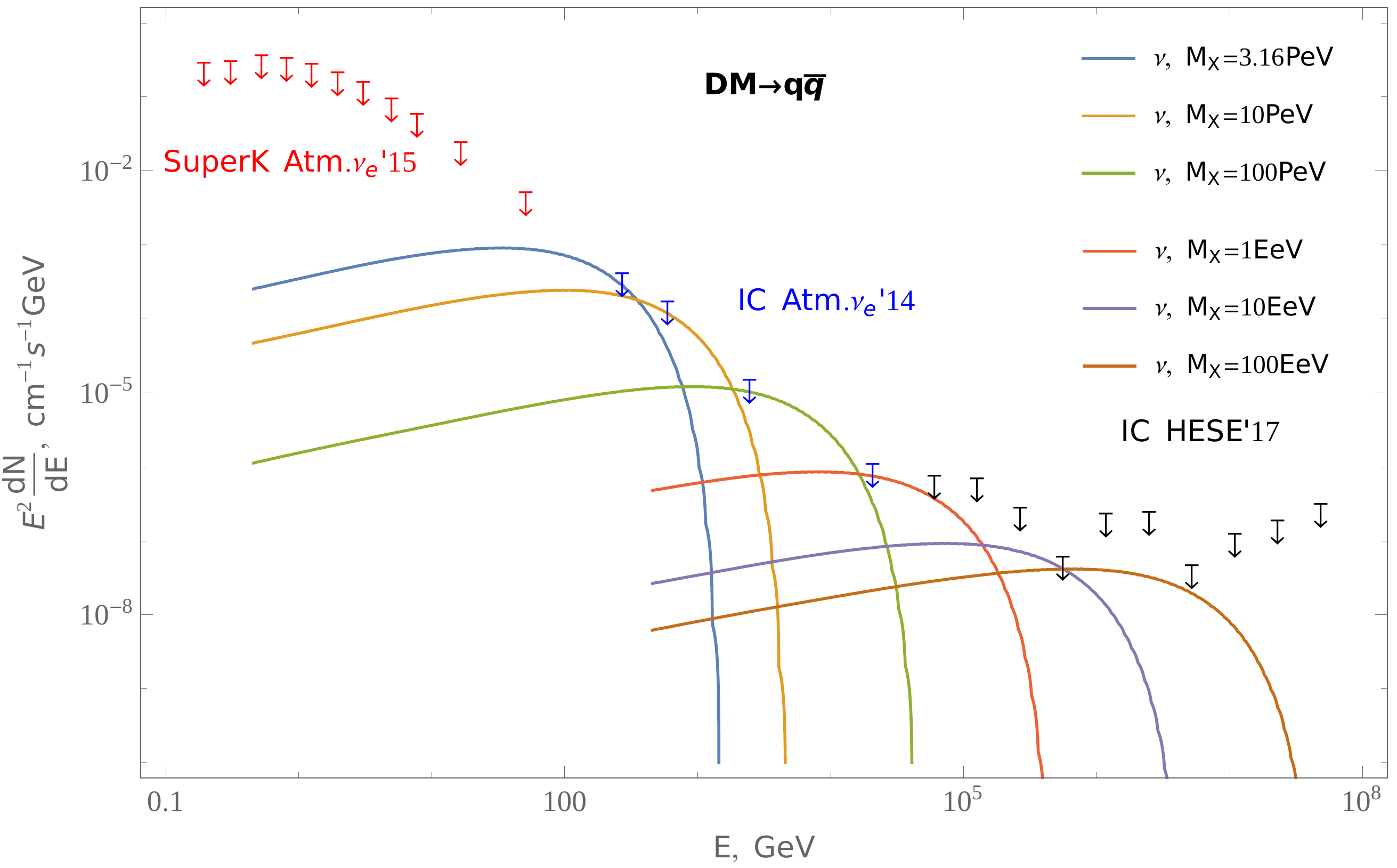}
	\figw{.49}{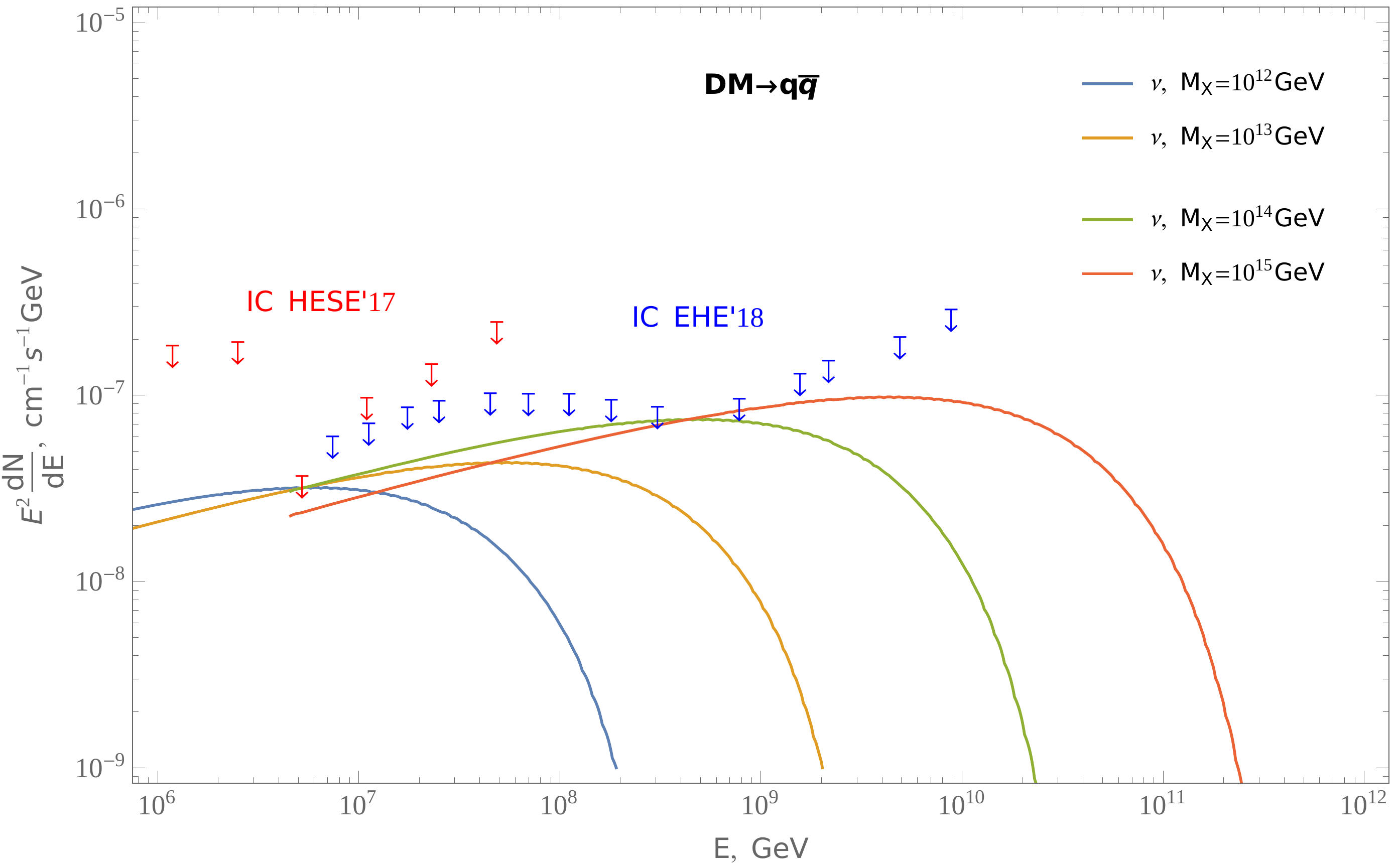}
	\figw{.49}{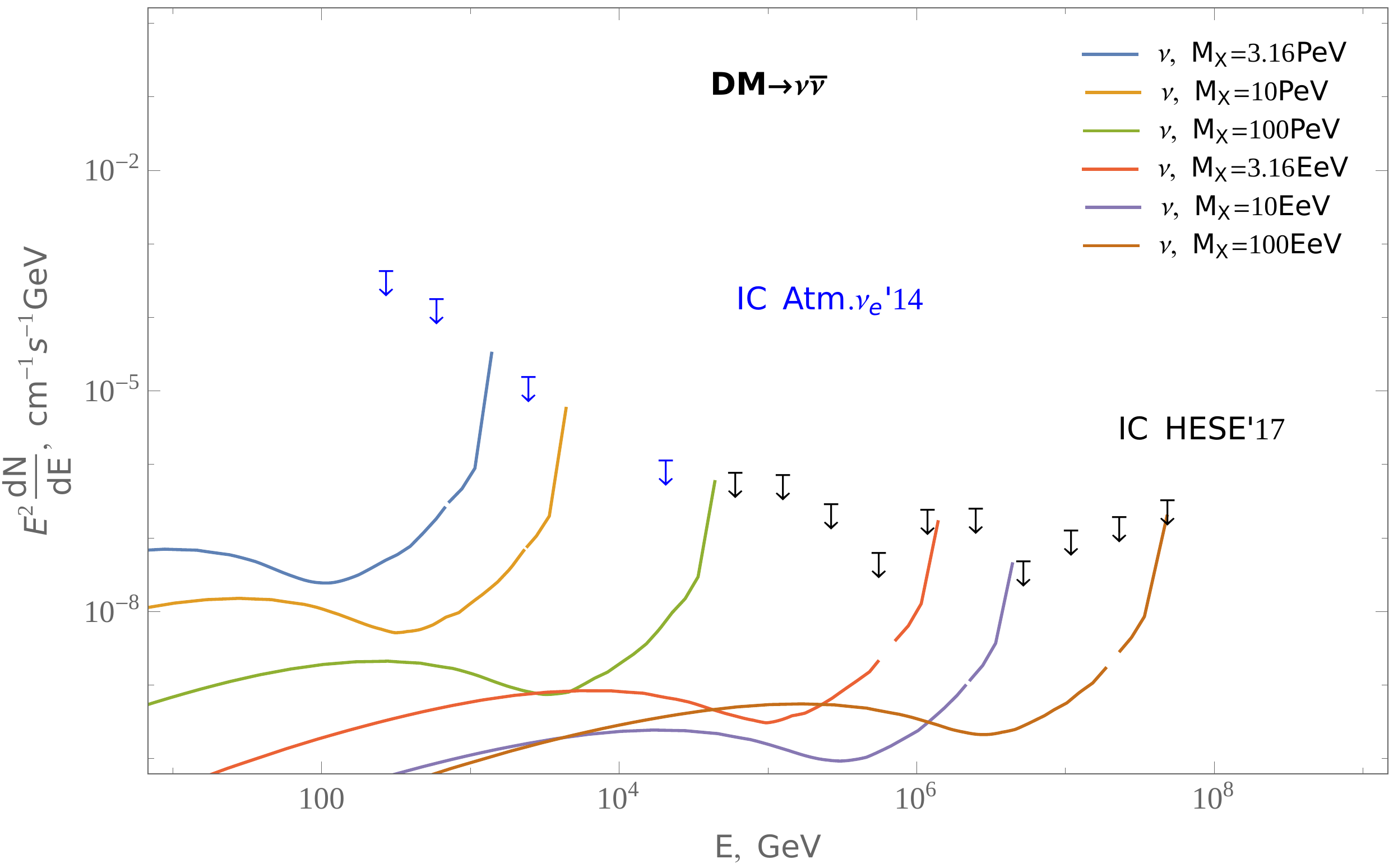}
	\figw{.49}{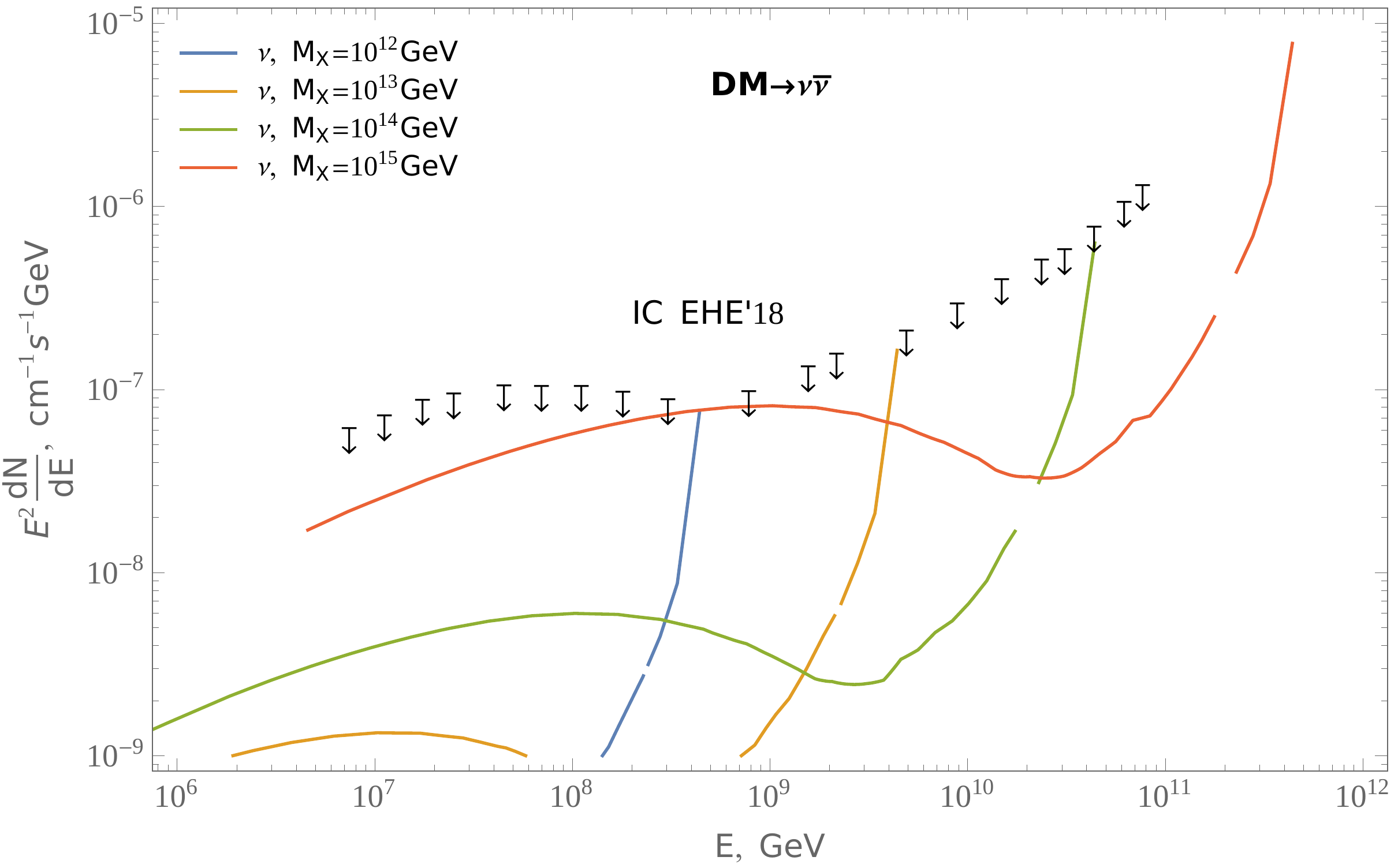}
	\caption{The spectra of $\nu$ from $q\bar{q}$ (top) and $\nu\bar{\nu}$ (bottom) decay
	for several DM particle masses higher than 1~PeV, compared with IceCube atmospheric electronic neutrino flux limits~\cite{Aartsen:2015xup} astrophysical (HESE set)~\cite{Aartsen:2017mau} and cosmogenic EHE set~\cite{Aartsen:2018vtx} \black neutrino flux limits.}
	\label{fig:spectrum_nu_hi}
\end{figure}

The independent set of constraints on the fraction of the DM decaying into visible particles could be derived using the recent experimental upper-limits on the diffuse neutrino flux in the wide energy range: from $\sim10$~MeV to $\sim100$~EeV. Namely, we adopt the Super-Kamiokande limits on the extragalactic supernovae neutrino~\cite{Zhang:2013tua}, limits derived from the IceCube data on the high-energy astrophysical neutrino (HESE set)~\cite{Aartsen:2017mau} and the IceCube limits on the highest-energy cosmogenic neutrino flux (EHE set)~\cite{Aartsen:2018vtx}. In the $100$~MeV -- $100$~TeV energy range the observed neutrino flux is dominated by the atmospheric neutrinos, that are difficult to discern from the cosmic ones. Therefore in this range we adopt experimental limits on the atmospheric neutrino flux derived by Super-Kamiokande~\cite{Richard:2015aua} and IceCube~\cite{Aartsen:2015xup} as a rough but conservative bounds for our model flux. 

Neutrino practically do not interact with medium during their propagation. Therefore we calculate their spectra at $z=0$ just by redshifting and assuming maximal mixing, i.e. $\nu_e:\nu_\mu:\nu_\tau = 1:1:1$ at $z=0$~\footnote{The resulting flavor composition for arbitrary initial flavour fractions was calculated in Ref.~\cite{Bustamante:2015waa}. The effect of deviations from $1:1:1$ composition does not exceed the overall systematic error in our approach.}. The neutrino injection spectra of $\nu\bar{\nu}$ and $e^+e^-$ decay channels contain sharp peaks near $E_\nu=M_X/2$. While the attenuation effects do not destroy the peaks, longer DM particle lifetime $\tau \gg H^{-1}$ may make them  smoother due to expansion of the Universe. Therefore, the constraints obtained in the assumption of short DDM lifetime would be stronger than those of long DDM lifetime, i.e. the short DDM lifetime assumption is not conservative in the case of neutrino constraints.

The examples of properly normalized neutrino spectra at $z=0$ for various decay channels and DM particle masses are shown in Fig.~\ref{fig:spectrum_nu_low} (low $M_X$) and in Fig.~\ref{fig:spectrum_nu_hi} (high $M_X$) in comparison with respective experimental neutrino flux limits.

\begin{figure}[!ht]
	\centering
	\figw{.49}{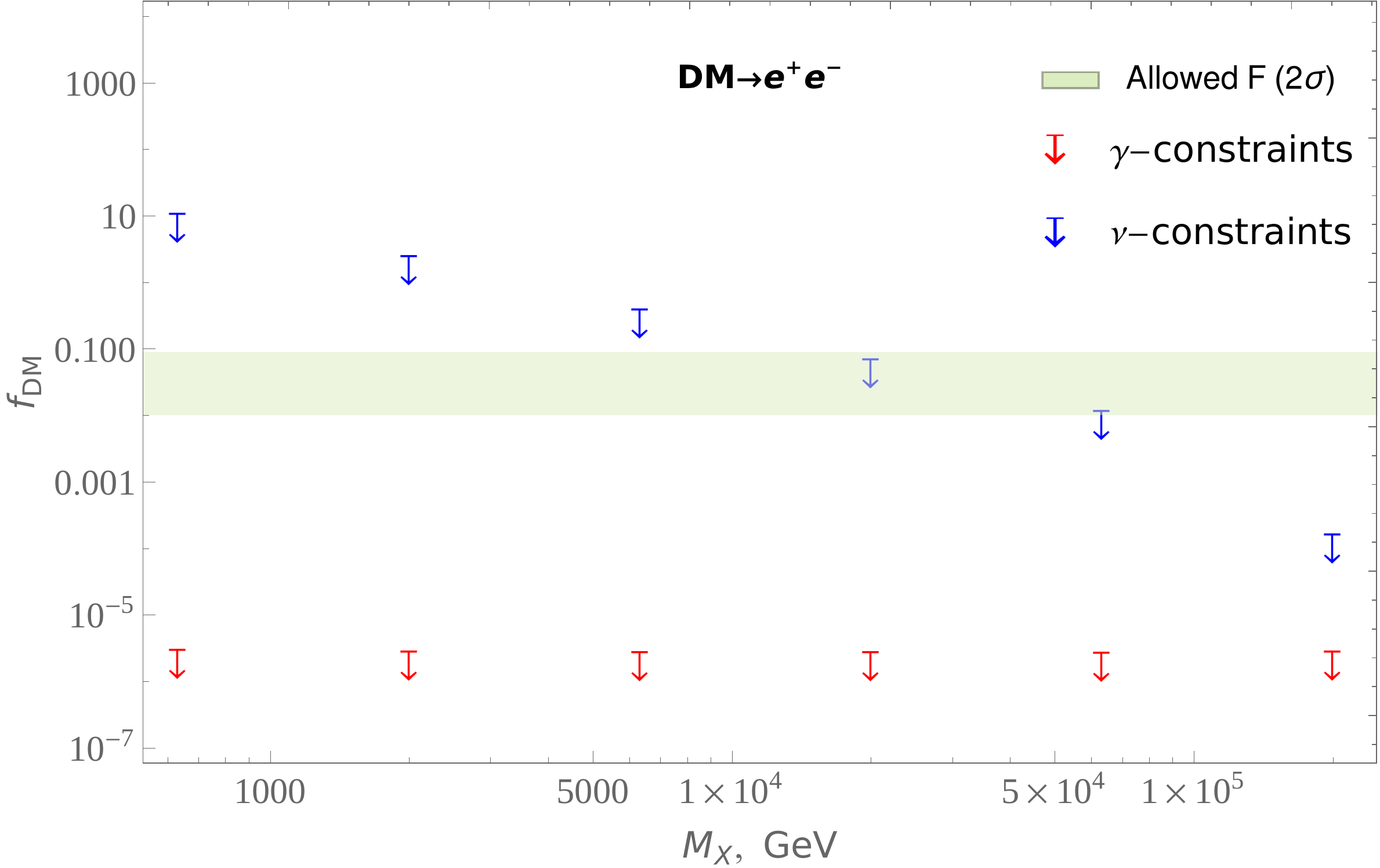}
	\figw{.49}{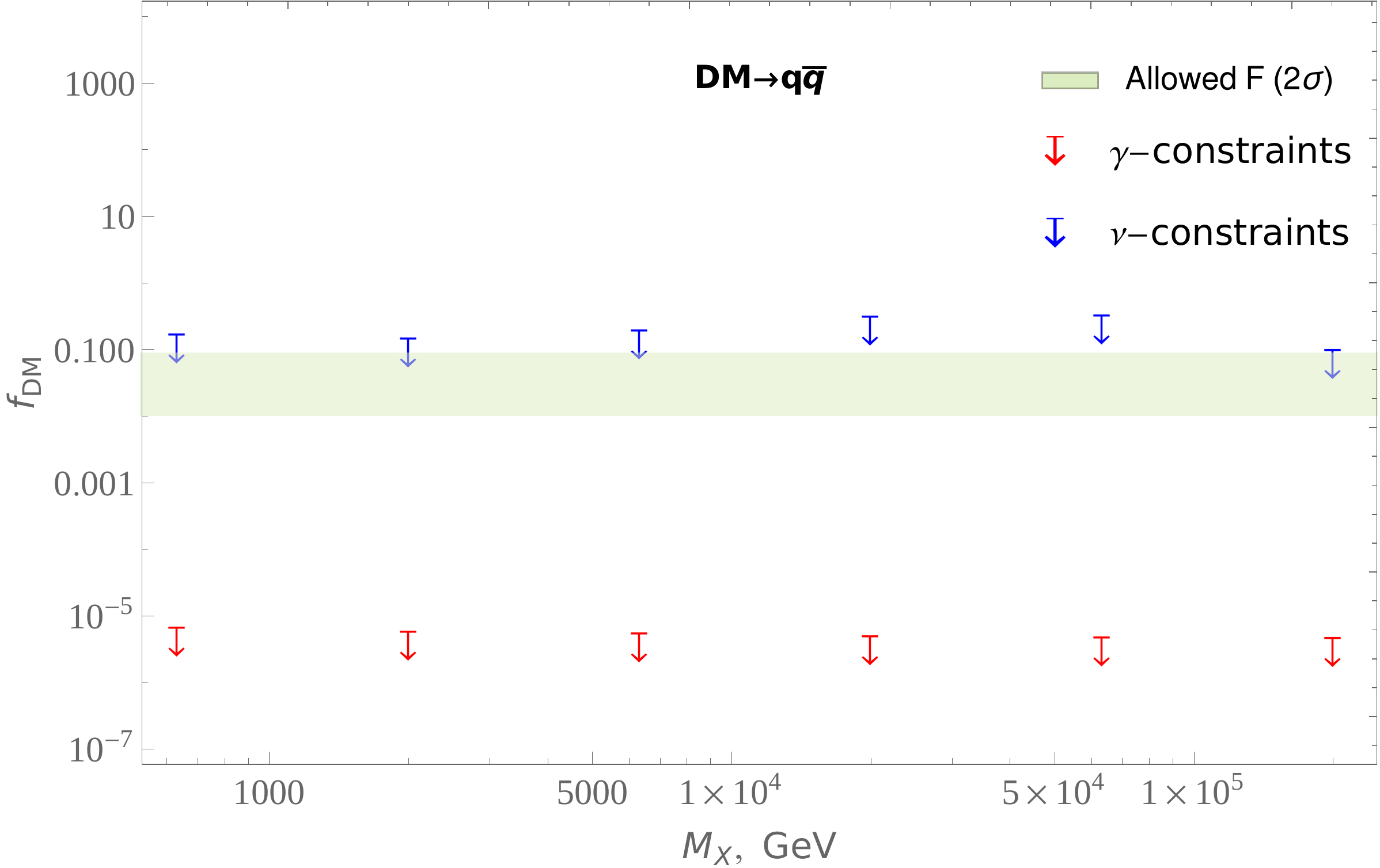}
    \figw{.49}{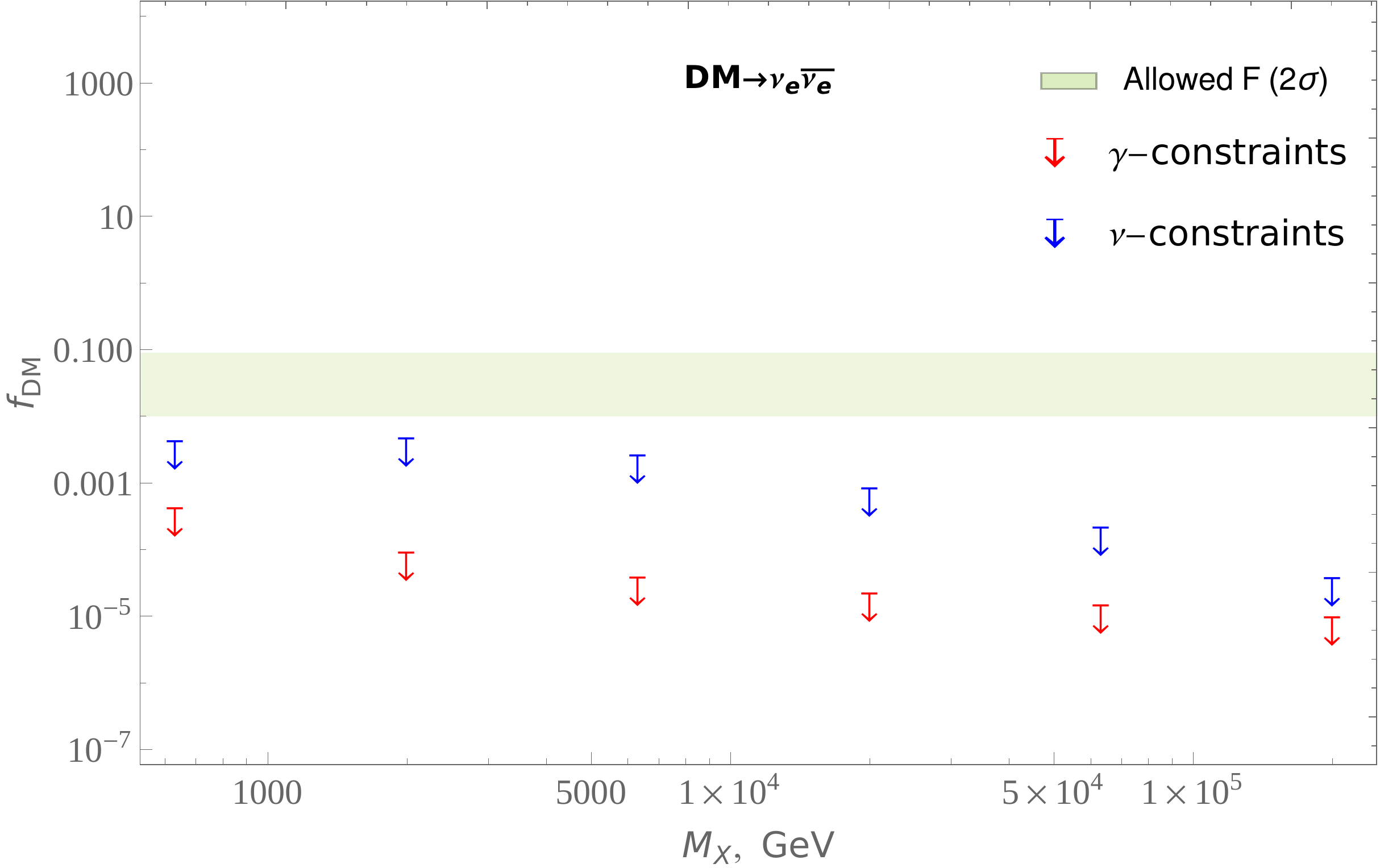}
	\figw{.49}{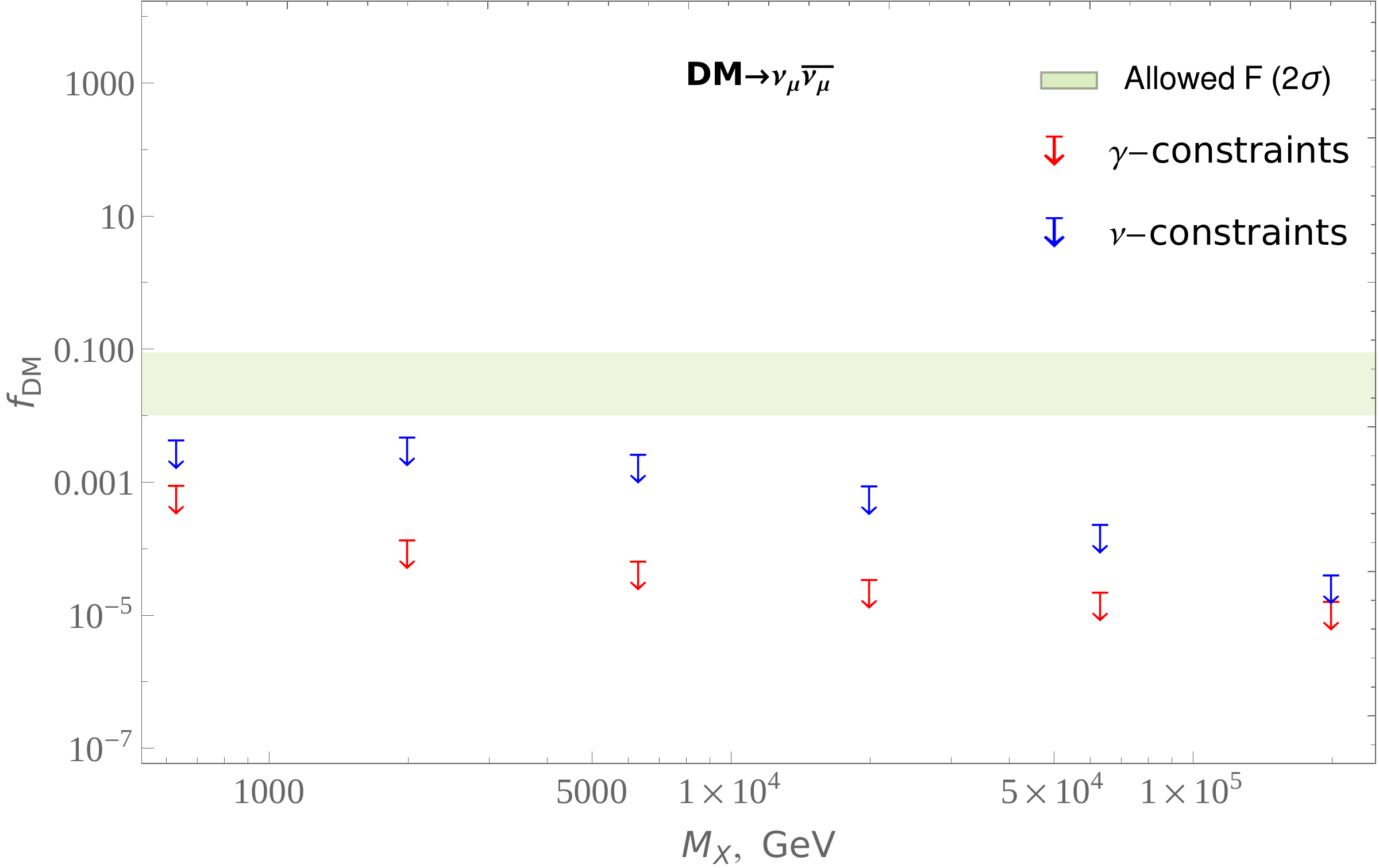}
	\figw{.49}{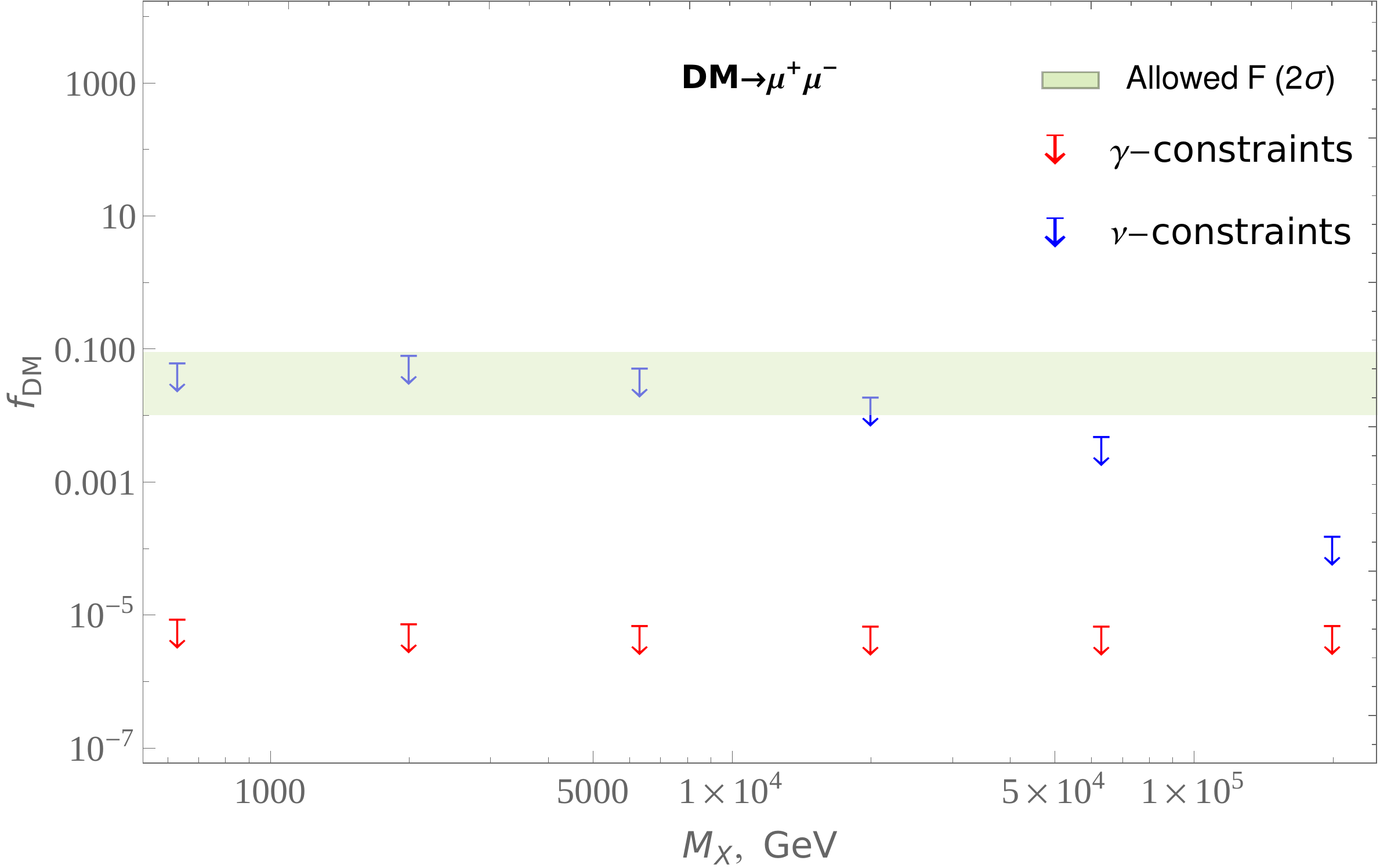}
	\figw{.49}{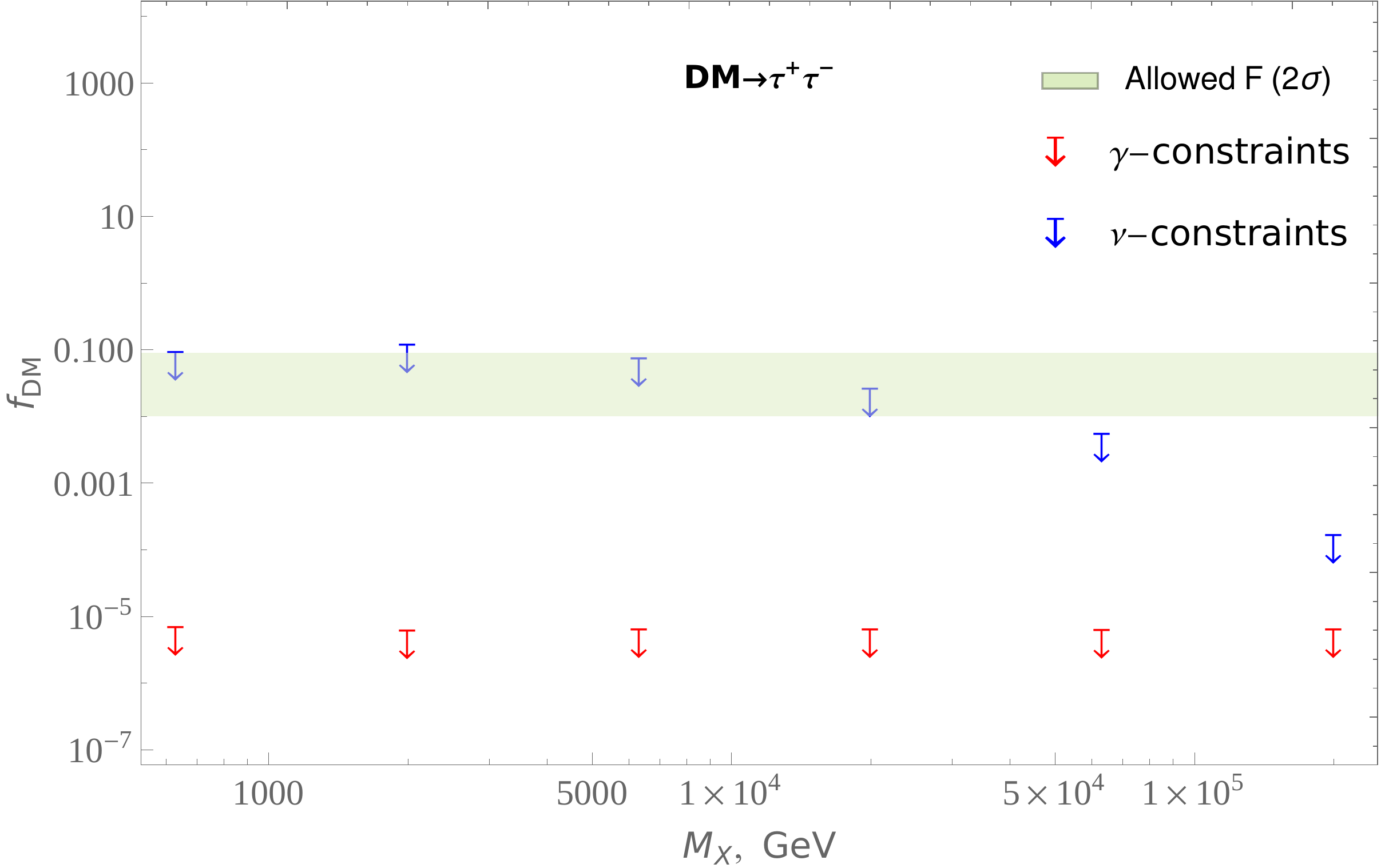}
	\figw{.49}{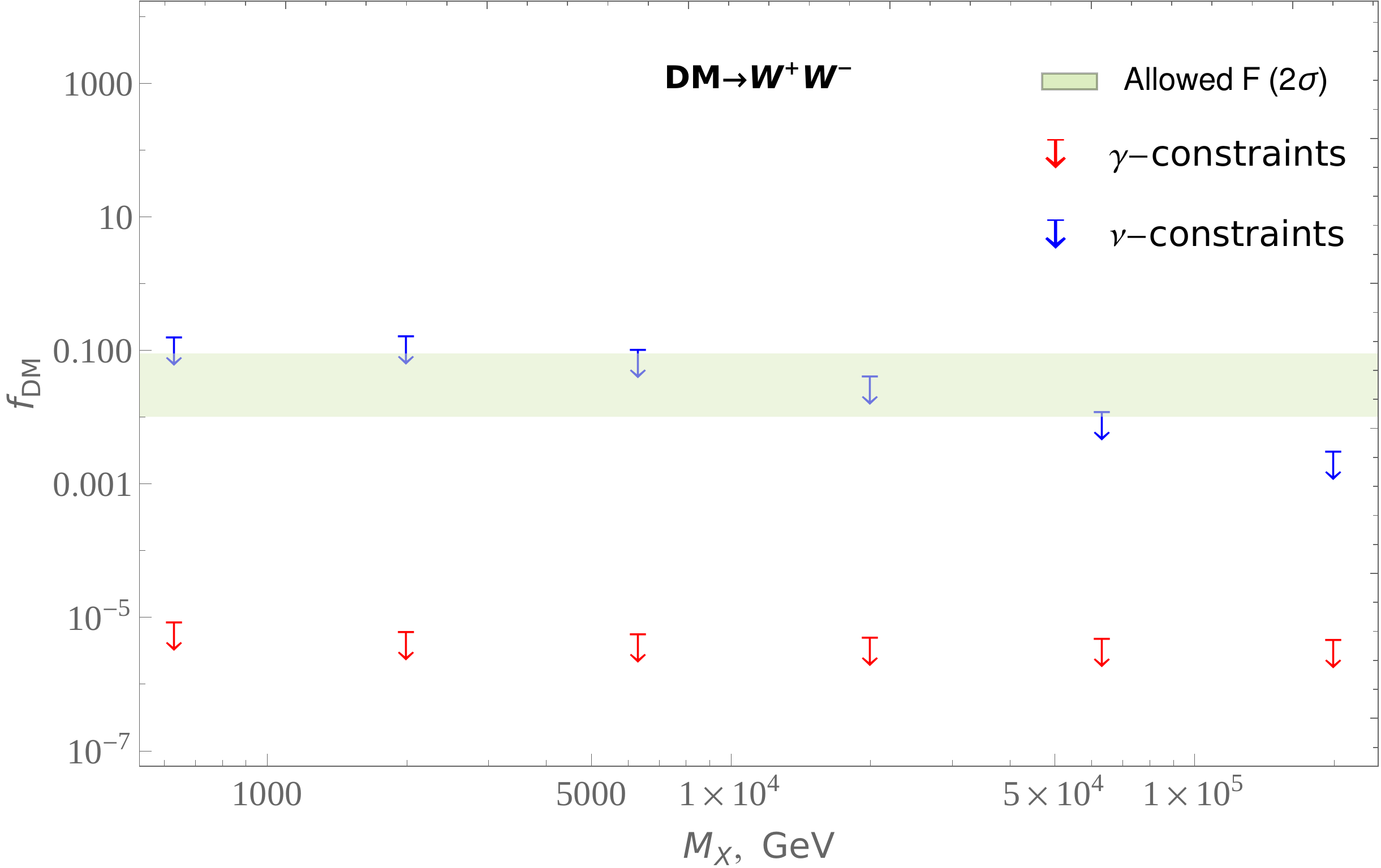}
	\figw{.49}{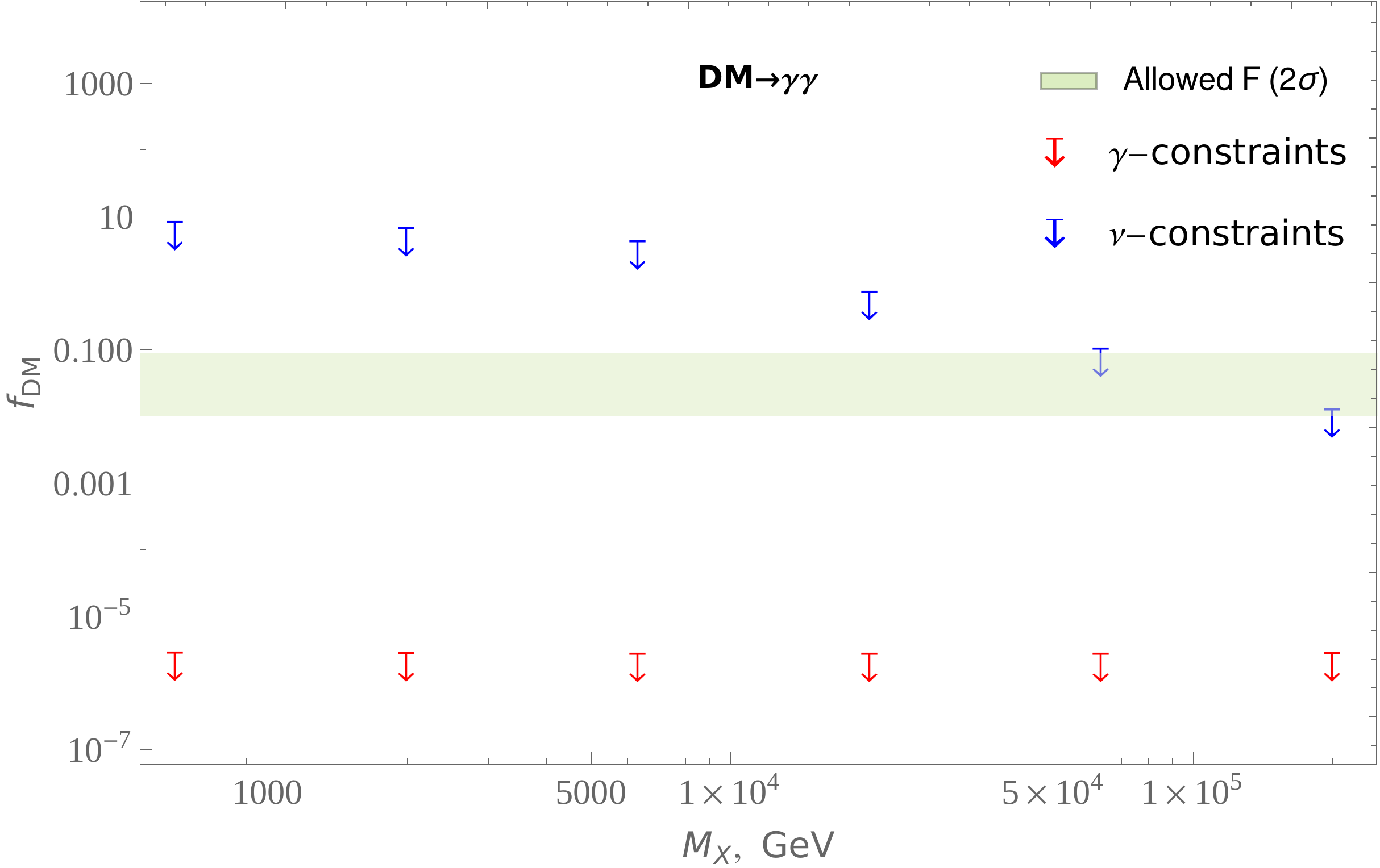}
	\caption{Constraints on a fraction $f_{\rm DM}$ of DM decaying into visible particles.
	Each plot shows the separate constraints from gamma-ray ($68\%$ C.L.) and from neutrino ($90\%$ C.L.) observations, for separate primary decay channel, left to right: $e^+e^-$, $q\bar{q}$, $\nu_e\bar{\nu_e}$, $\nu_\mu\bar{\nu_\mu}$, $\mu^+\mu^-$, $\tau^+\tau^-$, $W^+W^-$ and $\gamma\gamma$, for DM--particles of GeV-TeV masses. The $95\%$ C.L. constraints on a fraction $F$ of DM decaying after recombination, derived in Ref.~\cite{Chudaykin:2017ptd} from cosmological data analysis of Refs.~\cite{Zhao:2016das, Beutler:2011hx, Ross:2014qpa} is shown by the light-green band for comparison.}
	\label{fig:constr_low}
\end{figure}

\begin{figure}[!ht]
	\centering
	\figw{.49}{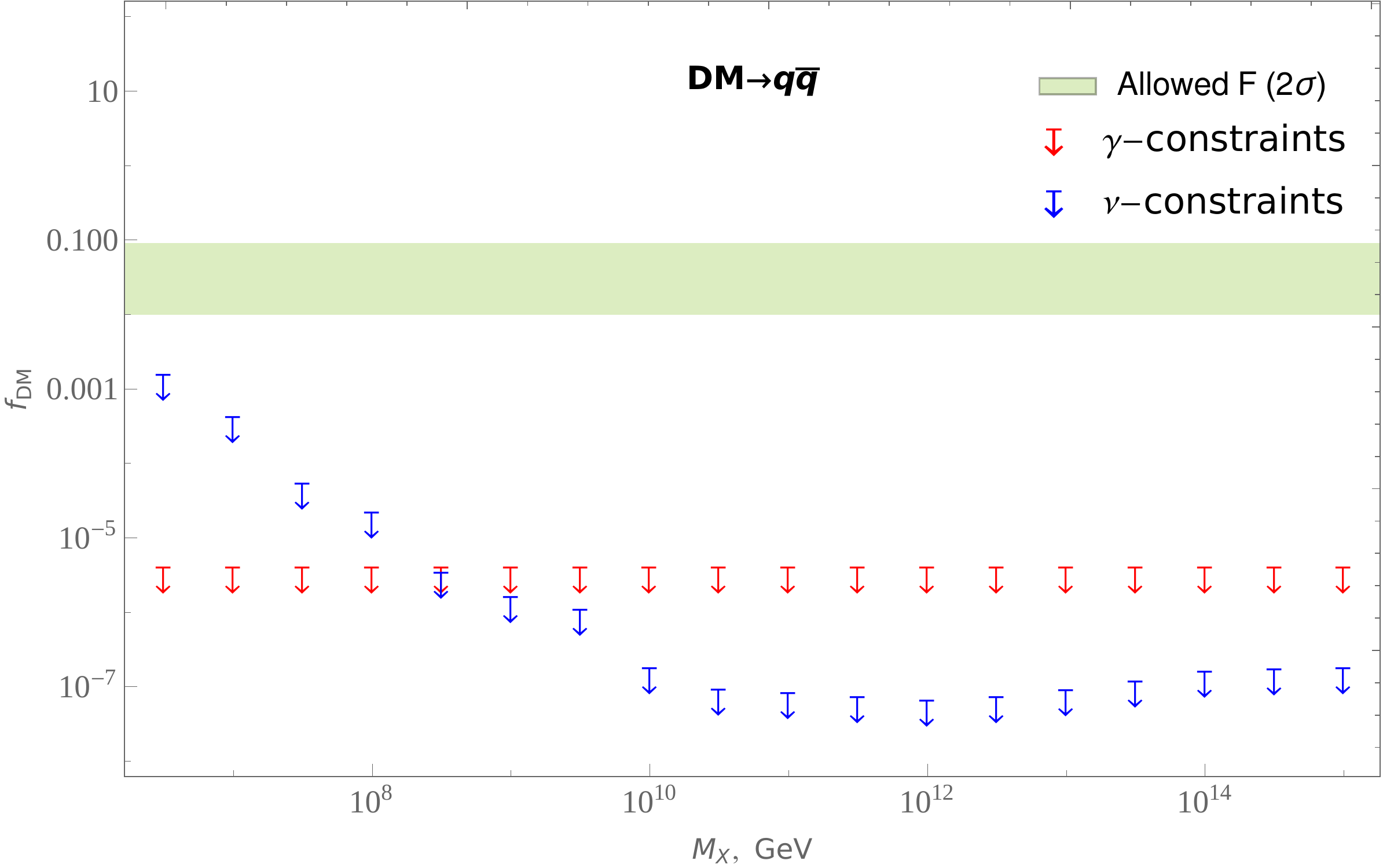}
	\figw{.49}{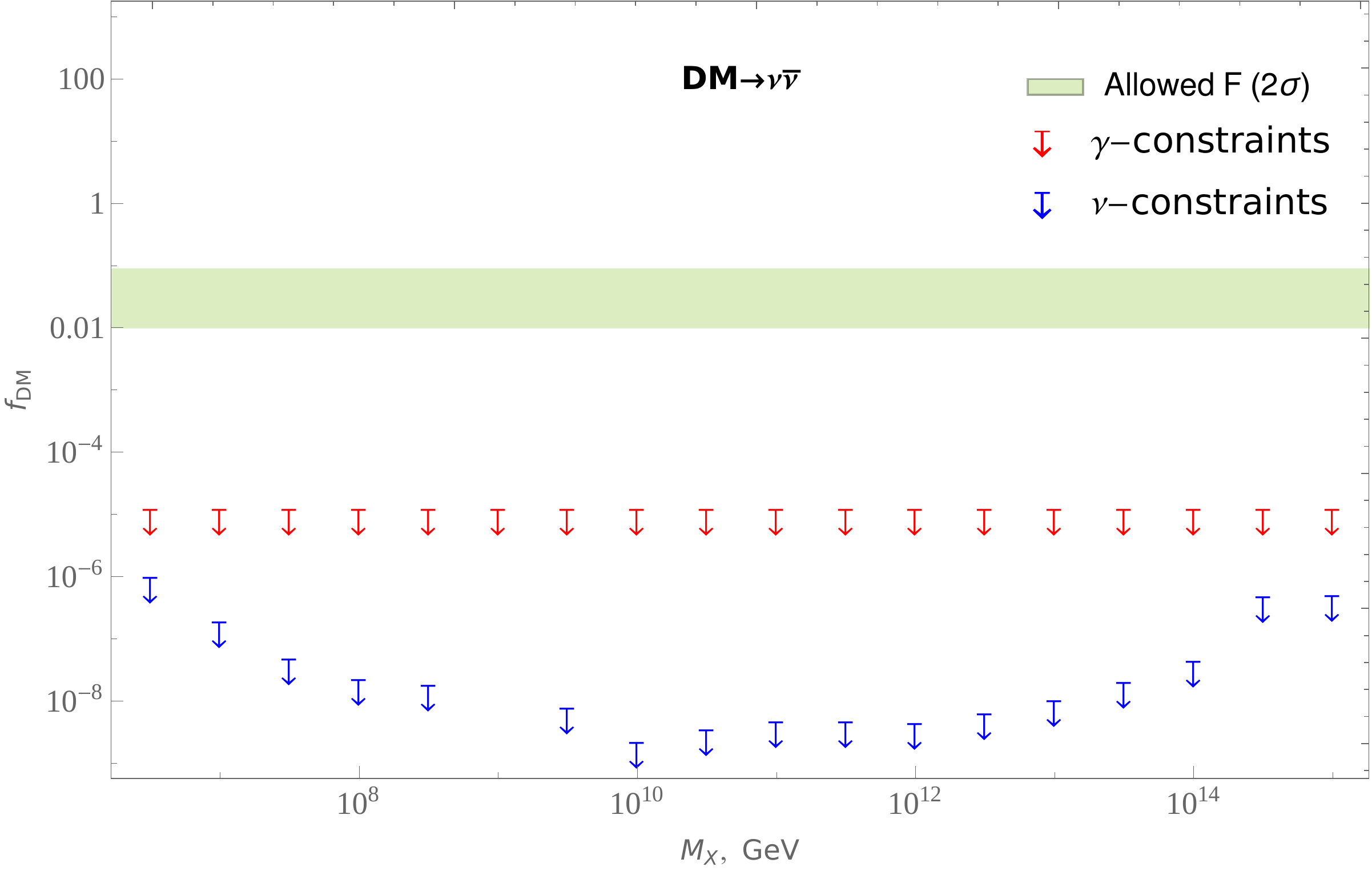}
	\caption{The same as Fig.~\ref{fig:constr_low}, but for DM--particles with PeV masses and higher and for two benchmark decay channels: $q\bar{q}$ (left), $\nu\bar{\nu}$ (right).}
	\label{fig:constr_hi}
\end{figure}

\section{Results}\label{sec:results}

In Fig.~\ref{fig:constr_low} we show constraints obtained on the fraction $f$ of the DM particles decaying into visible sector for $e^+e^-$, $q\bar{q}$, $\nu_e\bar{\nu_e}$, $\nu_\mu\bar{\nu_\mu}$, $\mu^+\mu^-$, $\tau^+\tau^-$, $W^+W^-$ and $\gamma\gamma$ decay channels for DM masses $M_X\leq 200$~TeV. Constraints derived with neutrino and $\gamma$-ray data are shown separately in blue and red points correspondingly. For comparison we also show in the same figures the preferred range of the decaying DM fraction ($0.05 \pm 0.04 (2 \sigma)$) derived in Ref.~\cite{Chudaykin:2017ptd} from cosmological data analysis of Refs.~\cite{Zhao:2016das, Beutler:2011hx, Ross:2014qpa}. One can see that for all the channels considered in the above energy range the $\gamma$-ray constraints are more strict allowing values of $f \lsim 10^{-5}$. The least strict limits were obtained not surprisingly in case of $\nu\bar{\nu}$ decay channels.

In Fig.~\ref{fig:constr_hi} we show the constraints obtained for larger masses $M_X\gsim$~PeV. The $\gamma$-ray constraints practically do not depend on $M_X$ since fraction of energy going to EM cascade is not changing with $M_X$, while the propagated $\gamma$-ray spectrum is universal. The constraints derived using neutrino data become more restricting than $\gamma$-ray constrains for $M_X \gsim 2\times 10^8$~GeV in case of  hadronic decay channel and for $M_X \gsim 10^6$~GeV in case of leptonic decay channel, reaching level of $f \lsim 3\times 10^{-9}$ for $M_X\simeq 10^{10}$~GeV in the latter case. These constraints are imposed by IceCube data. We conclude, that for all DDM mass range considered in this study, DM should decay mostly into invisible radiation, in order to match the $z=0$ $\gamma$-ray and neutrino measurements.

\acknowledgments
We would like to thank Igor Tkachev, Dmitry Gorbunov and Anton Chudaykin for helpful discussions.
The work was supported by the Foundation for the Advancement of Theoretical Physics and Mathematics “BASIS” grant 17-12-205-1

\newpage
\bibliographystyle{JHEP}
\bibliography{dm_iv}

\end{document}